\documentclass[times]{weauth}

\usepackage{moreverb}
\usepackage{listings}
\usepackage{graphicx}
\usepackage{float}
\usepackage{algorithmicx}
\usepackage[ruled]{algorithm}
\usepackage{algpseudocode}

\usepackage[hidelinks]{hyperref}

\newcommand\BibTeX{{\rmfamily B\kern-.05em \textsc{i\kern-.025em b}\kern-.08em
T\kern-.1667em\lower.7ex\hbox{E}\kern-.125emX}}
\algdef{SE}[DOWHILE]{Do}{doWhile}{\algorithmicdo}[1]{\algorithmicwhile\ #1}%

\begin{document}

\articletype{RESEARCH ARTICLE}

\title{A Web Scraping Methodology for Bypassing Twitter API Restrictions}

\author{ A. Hernandez-Suarez $^{1}$ ,G. Sanchez-Perez$^{1}$, K. Toscano-Medina$^{1}$, V. Martinez-Hernandez$^{1}$, V. Sanchez$^{2}$ and H. Perez-Meana $^{1}$}

\address{%
$^{1}$ \quad Instituto Politecnico Nacional, Graduate School ESIME Culhuacan; hmperezm@ipn.mx\\
$^{2}$ \quad University of Warwick, Department of Computer Science,  CV4 7AL, UK ; v.f.sanchez-silva@warwick.ac.uk}

\corraddr{Instituto Politecnico Nacional, Graduate School ESIME Culhuacan, San Francisco Culhuacan, CTM V, 04430 CDMX, Mexico; hmperezm@ipn.mx.}

\begin{abstract}
Retrieving information from social networks is the first and primordial step many data analysis fields such as Natural Language Processing, Sentiment Analysis and Machine Learning. Important data science tasks relay on historical data gathering for further predictive results. Most of the recent works use Twitter API, a public platform for collecting public streams of information, which allows querying chronological tweets for no more than three weeks old. In this paper, we present a new methodology for collecting historical tweets within any date range using web scraping techniques bypassing for Twitter API restrictions.
\end{abstract}


\maketitle

\section{Introduction}
Gathering proper information for training and testing data science algorithms is a primordial task that must be accomplished in order to obtain useful results. Many fields related to Natural Language Processing, Sentiment Analysis and Machine Learning use Online Social Network platforms to retrieve user information and transform it into machine-readable inputs, which are used by various algorithms to obtain predictive outputs like flu spreading detection \cite{lampos}, forecasting future marketing outcomes \cite{asur} and predicting political elections \cite{tuma}. Twitter is becoming the preferred  social network for data collection, in this network users can post short messages, also referred to as tweets, in a real-time manner; which has facilitated topic clustering research like rumour spreading analysis \cite{liakata}, human mobility sensing \cite{cuenca}, spam \& botnet detection \cite{haustein} and disaster response \cite{nandi}. The embedded information in a tweet may include images, geographic locations, url references and videos. Because of its usability and widespread, Twitter engines may dispatch approximately 1-billion of user-generated content per month, which is re-distributed over several countries around the world. 

Data researchers have started using Twitter for scientific approaches due to the facility for querying and collecting large volumes of data in a short time. As presented in early works like \cite{connor}, collecting tweets can be done by scraping streams of public available data by using Twitter "Gardenhouse" API \cite{twitter}, which is an endpoint designed for reading and writing tweets. 
After registering an application, the platform then returns a set of tokens granting  access to the streaming API. Well known works \cite{musta} \cite{conover} have used keywords for querying the Twitter API endpoint and retrieving formatted objects containing tweets with additional fields \cite{twitterd}; e.g., geographic coordinates, chronological information , retweet-metadata, favorite counts and language information. Although tweets are easily collected from the previously mentioned API, a big limitation occurs when stream rates exceed the number of retrieved tweets given $n$ number of queries. Twitter addresses this issue by limiting the flows of data to 15-minute interval by user application. In addition, another limitation is the fact that  
historical tweets are only available for a maximum range of three weeks by registering an enterprise API \cite{search-api} or by applying to a full-search access \cite{full-api} where potential clients are subjected to social and financial evaluations.   
Unfortunately, little is known about how to overcome this problem \cite{mckenna}. Some sites like Gnip\cite{gnip} and Sifter\cite{sifter} offer a full archive of old tweets by registering and paying for volumes of tweets, sometimes at expensive prices.  In this paper, we propose a web scraping methodology for crawling \cite{khalil} and parsing tweets bypassing Twitter API restrictions taking advantage of public search endpoints , such that,  given a query with optional parameters and set of $HTTP$ headers we can request an advanced search going deeper in collecting data.

\section{A Web Scraping Solution}

\begin{figure}[h]
\centering
\includegraphics[width=12 cm]{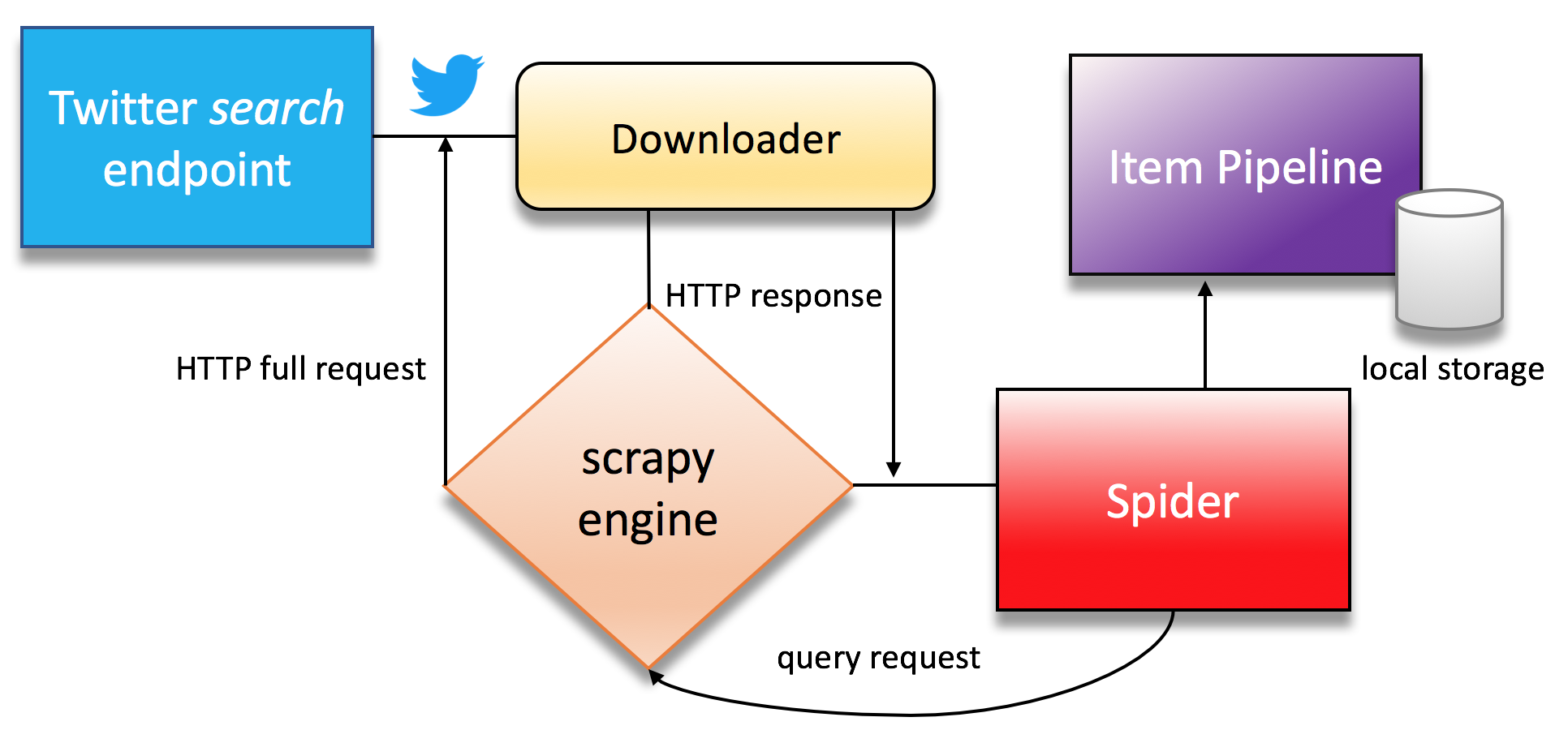}
\caption{Proposed scrapping scheme.}
\label{1}
\end{figure}

Web scraping techniques are used to extract information from websites in an automatic way \cite{suganya} by parsing hypertext tags and retrieving plain text information embedded onto them. We propose a new methodology for scraping Twitter Search endpoint and customizing queries fields in order to extend searching capabilities. At a glance, web scrapping seems an easy task but analyzing HTTP requests and responses \cite{singh} can be really complex. By using Scrapy, an \textit{open source and collaborative framework for extracting data from websites} \cite{scrapy} written in Python, we enhance the power of scraping engines to obtain an unlimited volume of tweets bypassing date ranges limitations. Our proposal is graphically depicted in Figure \ref{1}.

The work flow is described next. Given any valid set of \textit{HTTP-HEADERS}, an \textit{array of words} (also known as querying  terms) , an array of dates (date range) and \textit{optional parameters}, a query request is sent to the Twitter Search endpoint, composing then an URL(\ref{one})

\begin{multline}
https://twitter.com/search?f=tweets\&vertical=default\&q=
\textbf{words}+\textbf{array of dates}+\\\textbf{parameters}
\label{one}
\end{multline}

A rendered HTTP response is then processed by a web spider and the HTML payload is redirected to a download layer. Finally,  unprocessed data containing tweets is fed into the  Scrapy engine in order to strip hypertext tags by  objects known as $tag-selectors$; each tweet is treated as an independent structure composed of plain text, date of creation and geographic information (if exists). When scraped, responses are retrieved with a $maximum-position$  class attribute from an $<div>$ HTML tag; this information is a pagination identifier from the last appended tweets. The reason for having this attribute is because if we want to scrap deeper tweets, the  $maximum-position$ information must be fed into a second search endpoint, as shown in URL (\ref{two})

\begin{multline}
https://twitter.com/i/search/timeline?f=tweets\&vertical=default\&q=\textbf{words}\\+
\textbf{array of dates}+
\textbf{parameters}
\&src=typd  
\&min\_position=\textbf{maximum-position}
\label{two}
\end{multline}

Approximately, 20 tweets can be collected in a single request, but those sent to the second search endpoint can scrap $k$ rounds of  asynchronous responses containing blocks of tweets. Unlike the first search endpoint, the second one is designed for users scrolling the main Twitter search feed and appending older tweets retrieved in JSON (JavaScript Object Notation) format, which are  internally processed by Twitter scripts. Collected tweets by Scrapy crawlers are processed through  a pipeline configured to store each one as an item in a relational or non-relational database client. The arguments for an advanced query based in \cite{twitters} are listed in Table \ref{table1}, the code for running the scraper with parameters on command line is depicted in Listing \ref{bash}:

	\begin{lstlisting}[language=bash,caption={Bash version},label={bash},xleftmargin=-1.0cm]
	#!/bin/bash
	$ scrapy crawl twitter -a [K-ROUNDS ] -a [ALL-WORDS] 
	-a [DATE-RANGE] -a [PARAMETERS]
	\end{lstlisting}
	
\begin{table}[H]
\caption{Arguments for an Advanced Search Query.}
\centering
\begin{tabular}{ccc}
\toprule
\textbf{Argument}	& \textbf{Parameter}	& \textbf{Description}\\
\midrule
 w & - -all-words=ARRAY OF WORDS & All words listed in an array \\
  k & - -k-rounds=INTEGER & Number of rounds for retrieving
  deeper tweets  (1 by default)  \\
  e  & exact-phrase=STRING & An exact phrase\\
  aW &  - - any-words=ARRAY OF WORDS & Any words listed in an array \\
  h & - -hashtag=ARRAY OF WORDS & Any words listed in an array\\
  l &  - -language=ISO CODE & Any ISO code for a given language\\
  p & - -account=STRING & Word(s) or Hashtag(s) coming from an specified account\\
  pF & - -account-from=STRING & Word(s) or Hashtag(s) going from an specified
   account\\
   pM & - -account-mention=STRING & Word(s) or Hashtag(s) mentioning an specified
   account\\
   g & - -near-place=STRING & Word(s) or Hashtag(s) published inan specific location\\
   gK & - -near-place-miles=INTEGER & Number of miles radio (15 miles by default)\\
   d & date-range=ARRAY OF DATES & Date range (the first element of date array is $since$, last is $until$)\\
\bottomrule
\end{tabular}
   \label{table1}
\end{table}

Our proposed methodology is described in Algorithm \ref{array-sum}

\begin{algorithm}
\caption{Extracting $n$  number of tweets using a web scraping methodology}
\label{array-sum}
\begin{algorithmic}[1]
\Procedure{TwitterScraping}{$k_{rounds},words,dates,parameters$}\newline
\textbf{Input: $endpoint_{1}, endpoint_{2}, maximum_{position}=0$, $tweets = \{\}$} \newline
\textbf{Output: $tweets= tweet_i\in\{tweet_{text},tweet_{date},tweet_{geodata}\}_{i=1}^n$} \newline
    \For {each round $i=0$ to $k_{rounds}$}
    		 \For {each date in  $dates$}
    		\State $pagination \leftarrow 0$
    		  \Do
    		 	\State $scrap(words,date,endpoint_{1},parameters) = tweet$
    		 	\State Append each $tweet$ to $tweets$
    		 	\State Extract current $pagination$  from $tweet$
    		  \doWhile{$ scrap(w,d,endpoint_{1},p) \neq NULL$}
    		  \State $ maximum_{position} \leftarrow pagination $
    		   \Do
    		 	\State $scrap(words,date,endpoint_{2},parameters) = tweet_{maximum_{position}}$
    		 	\State Append each $ tweet_{maximum_{position}}$ to $tweets$
    		  \doWhile{$ scrap(words,date,endpoint_{2},parameters) \neq NULL$}
    		 \EndFor
	\EndFor
	\State Return $tweets$
\EndProcedure
\end{algorithmic}
\end{algorithm}

\section{Deploying and Deamonizing the Scraper}

\begin{figure}[H]
\centering
\includegraphics[width=12 cm]{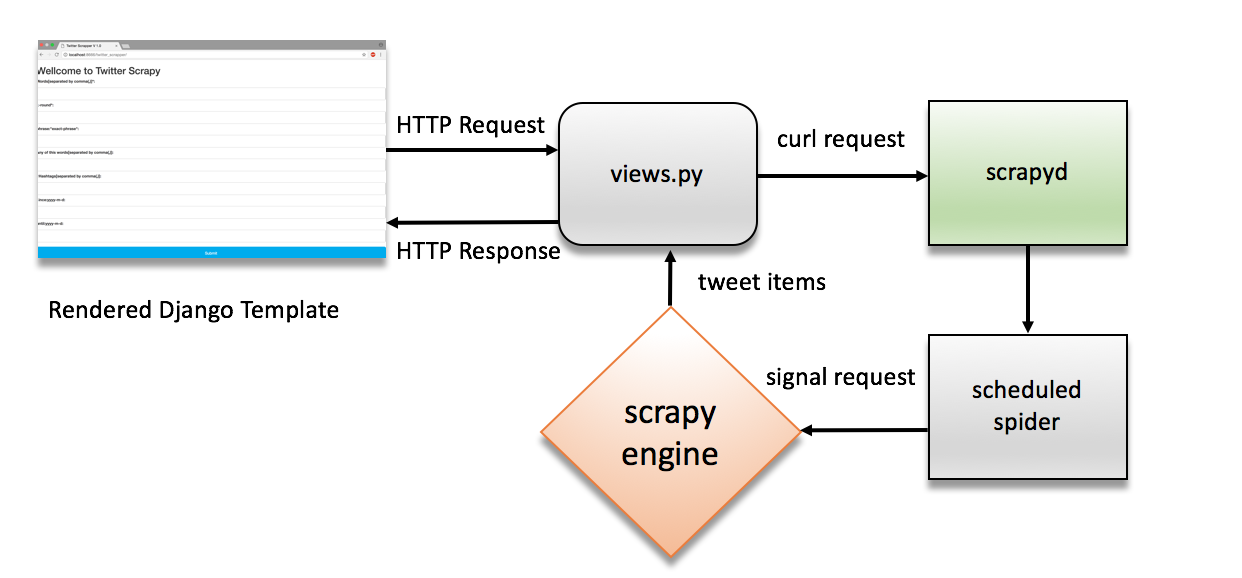}
\caption{Daemonizing Scrapy work flow.}
\label{2}
\end{figure}

In most applications, it is important to consider  usability \cite{jones}; because of the nature of Python applications, scripts are run via  command line  with required arguments and the binary code is then launched to perform programmed operations.  In this work, we build a Graphical User Interface (GUI) for testing our scraping approach. Specifically we develop a web service with Django\cite{django}, a framework for developing web applications using Python as base language,  
which includes a model-view-controller design for quickly project escalation. Scrapy spiders cannot start instances for web crawling ; this is  because each task is triggered as an independent process and due to Django security restrictions (regarding  operating system privileges) all non-root processes are denied. To solve this issue, we bind the Django GUI to \textit{scrapyd} \cite{scrapyd}, a service that \textit{damenonize} crawling tasks. Similarly to unix-like \textit{cron} (job scheduler) tasks, Scrapy spiders can be scheduled to crawl targeted websites by calling scrapyd background instances using \textit{curl} commands as shown in Listing \ref{bash2}.:

	\begin{lstlisting}[language=bash,caption={bash command for scheduling scrapyd job},label={bash2},xleftmargin=-1.0cm]
	#!/bin/bash
	$ curl http://localhost:6800/schedule.json -d project=twitter
	_spider -d spider=twitter -d [PARAMETERS]
	\end{lstlisting}

Django \textit{view layer} contains handlers for HTTP requests and responses from user activity performed on the GUI  template. When a request is received, a signal is sent to launch the scraper, thus  binding a scheduled job deployed by scrapyd to crawl  the previously mentioned Twitter search endpoints.  The work flow for daemonizing Scrapy is depicted in Figure \ref{2}.

\begin{figure}[H]
\centering
\includegraphics[width=12 cm]{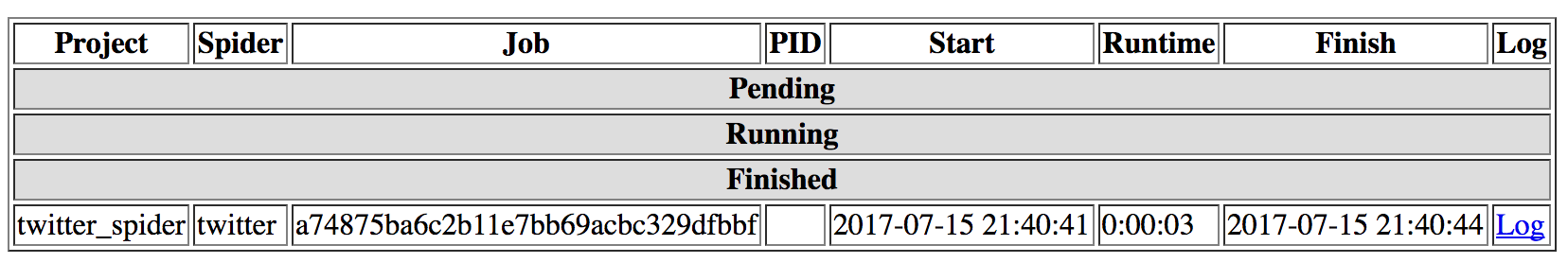}
\caption{Scrapyd scheduled jobs interface.}
\label{3}
\end{figure}

Scrapyd daemon contains also an interface for monitoring stacked jobs launched from Django view layer. Figure \ref{3} depicts the scheduled jobs interface.

\begin{figure}[H]
\centering
\includegraphics[width=14 cm]{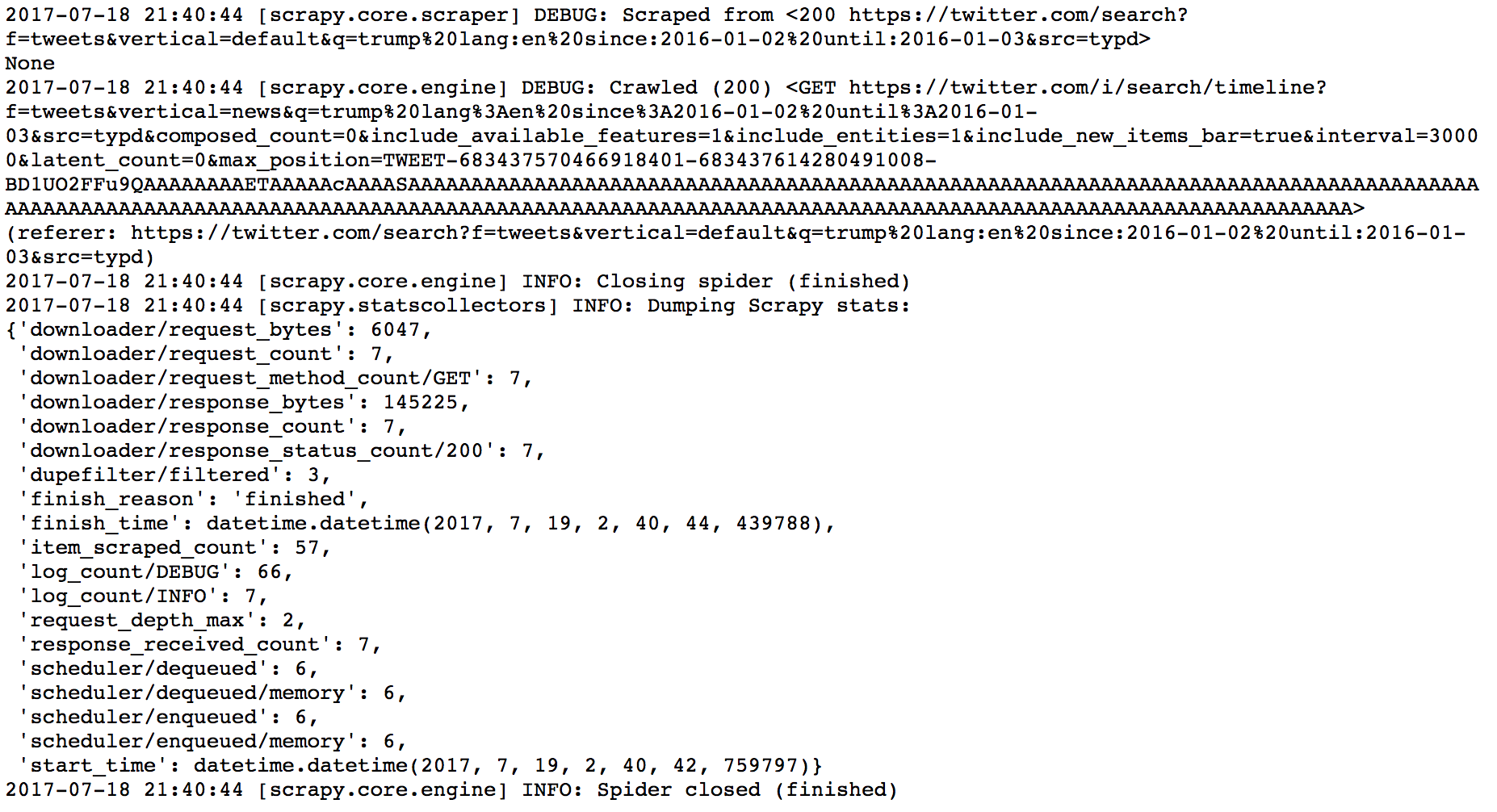}
\caption{Information Scraped from Twitter Search Endpoints.}
\label{4}
\end{figure}

Example of values for building a single query to scrap tweets are shown in Table \ref{table2}, a log describing crawled urls and debugging information from 
Twitter Search Endpoints is depicted in Figure \ref{4}

\begin{table}[H]
\caption{Values for building a single query.}
\centering
\begin{tabular}{ccc}
\toprule
\textbf{K-ROUNDS}	& \textbf{DATE-RANGE}	& \textbf{ALL-WORDS}\\
\midrule
 4 & 2016-1-2 to 2016-1-2  & trump\\
\bottomrule
\end{tabular}
   \label{table2}
\end{table}

\begin{figure}[H]
\centering
\includegraphics[width=10cm]{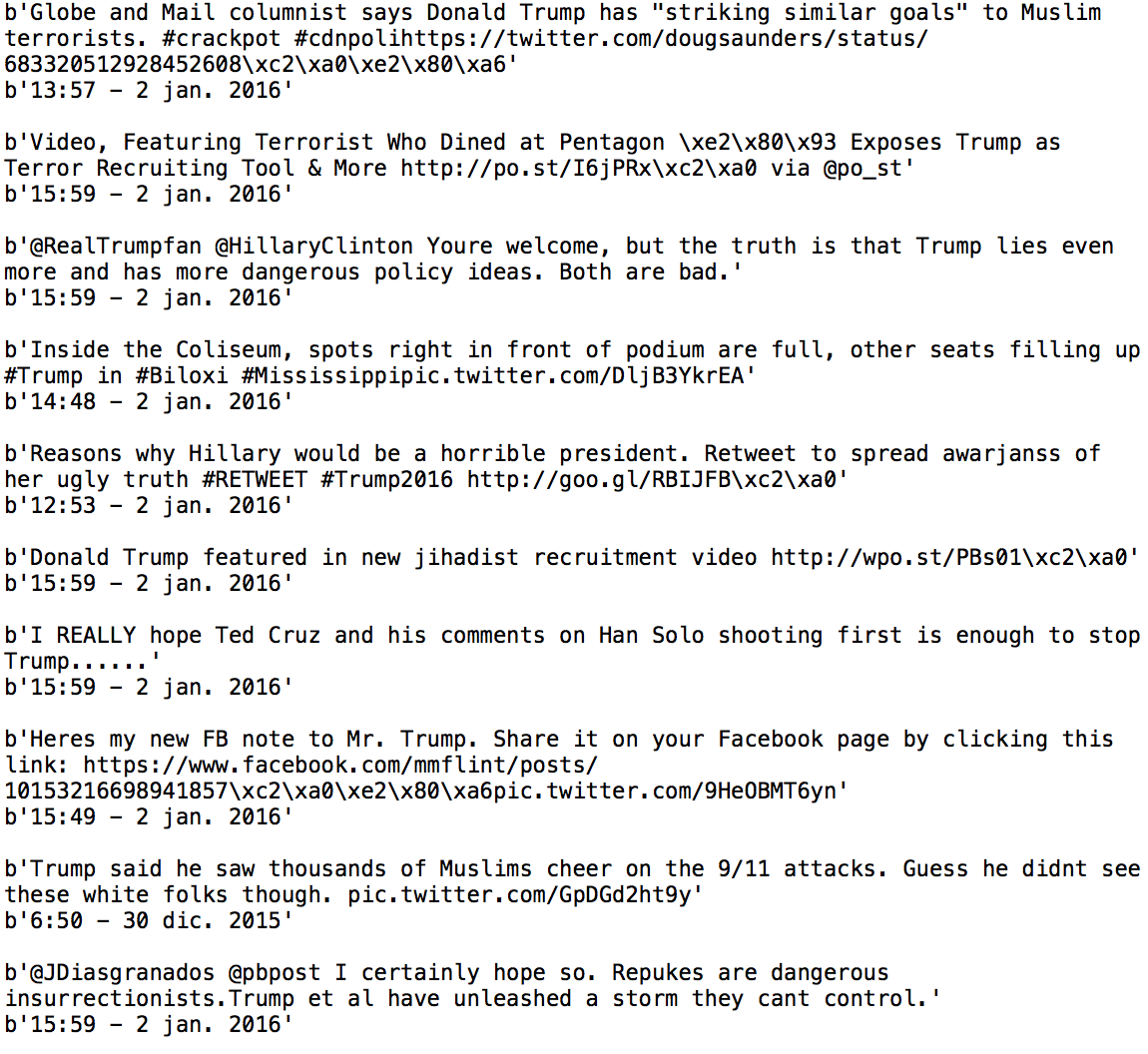}
\caption{Excerpt of retrieved tweets.}
\label{5}
\end{figure}

Responses from the Scrapy engine are retrieved as instances from a Scrapy class named $Items$,  but they can be transformed into comma-separated values or plain text files. An example of plaint text tweets scraped from the Twitter search endpoints based on the query in Table \ref{table2} are depicted in  Fig. \ref{5}.

\section{Evaluating the Scraper Performance}

\begin{figure}[H]
	\centering
	\includegraphics[width=16cm]{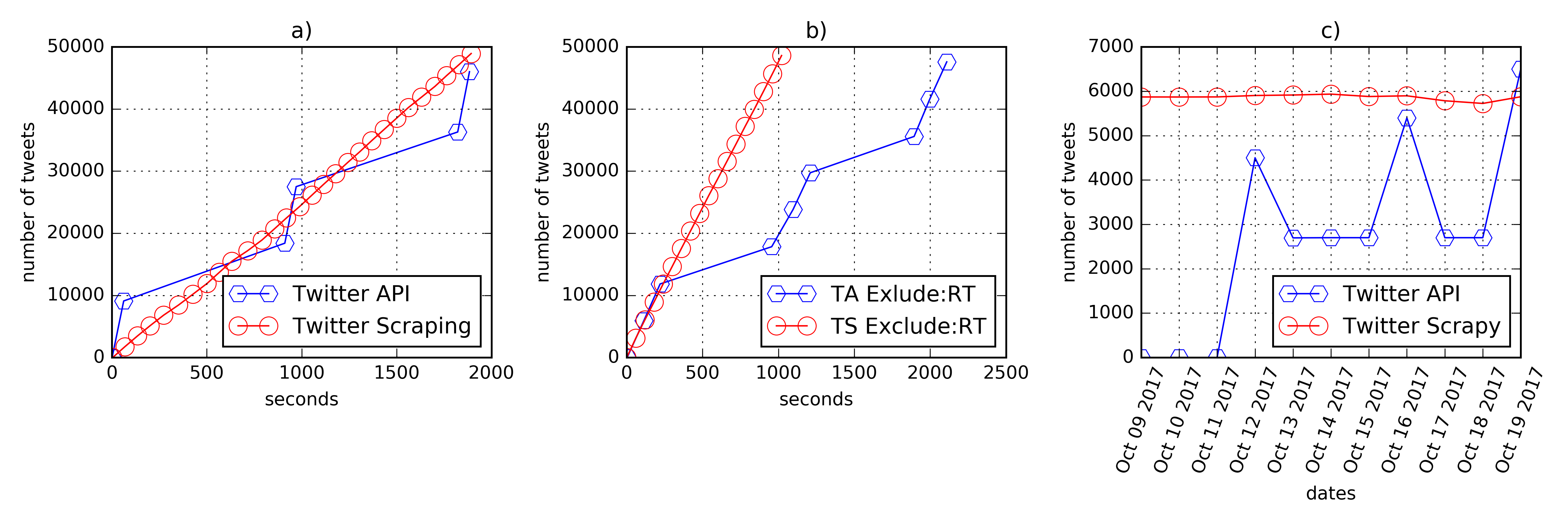}
	\caption{Comparing the performance between Twitter API and Scrapy}
	\label{6}
\end{figure}

Evaluating scraping performance is useful to contrast Twitter Garden house API (Stream \& Search) and our proposed methodology (Twitter Scrapy). The set of metrics for comparing both approaches are : the \textit{total amount of time} for retrieving blocks of tweets, the \textit{volume of tweets} retrieved for a query $q$ and the
 maximum number of historical tweets given a range of dates. 
The number of retrieved tweets by querying the stream without any filters is plot in  Fig \ref{6}a;  Fig \ref{6}b plots the number of retrieved tweets by filtering original statuses (no retweets and mentions) and finally \ref{6}c plots the number of retrieved tweets for a span of one week.

In Table \ref{table3}. is compared the results of the proposed metrics of Twitter API and Twitter  Scrapy  methodology 

\begin{table}[H]
\caption{Results of comparing Twitter API and Twitter Scrapy metrics. Highlighted values are those who improved time and volume for retrieving tweets.}
\centering
\begin{tabular}{cccc}
\toprule
\textbf{Methodology}	& \textbf{DATE-RANGE}	& \textbf{Total of  retrieved tweets} & \textbf{Seconds}\\
\midrule
 Twitter API Stream & - & 46000 & 1800\\
  \textbf{Twitter Scrapy} & -  &  \textbf{48929} & 1893 \\
 Twitter API Search Excluding Retweets and Mentions & - & 47582 & 2109\\
  \textbf{Twitter Scrapy Excluding Retweets and Mentions} & - & \textbf{48601} & \textbf{1021}\\
  Twitter API Search & 9-10-20107 \textit{to} 19-10-2017 & 29895 & - \\
  \textbf{Twitter  Scrapy} & \textbf{9-10-20107} \textit{to} \textbf{19-10-2017} & \textbf{64531} & - \\

\bottomrule
\end{tabular}
   \label{table3}
\end{table}

\section{Conclusions}

 Gathering information from Online Social Networks is a primordial step in many data science fields allowing researchers to work with different and more detailed datasets. Although an important proportion of the scientific community uses the Twitter streaming API for collecting data, a limitation occurs when queries exceed rating intervals and time ranges. 
In this article, we presented a new methodology for querying Twitter Search Endpoints bypassing their "gardenhouse" API restrictions, in order to retrieve large volumes of data generated over longer periods of time with faster results. 
We also showed that by using Python technologies such as Scrapy, Django and customized daemons, it is possible to develop and escalate a web interface for launching, controlling and retrieving information from web crawlers. Our development can be tested in an Amazon EC2 Web Services cloud environment. The application can be accessed publicly at \url{http://ec2-13-58-43-111.us-east-2.compute.amazonaws.com:8001/twitter_scrapper/} or by an ip address \url{http://13.58.43.111:8001/twitter_scrapper/}.

\section*{Conflict of Interest Disclosure }

The author(s) declare(s) that there is no conflict of interest regarding the publication of this paper

\end{document}